\documentclass[12pt,letterpaper]{article}
\usepackage[top=0.85in,left=0.85in,footskip=0.75in,marginparwidth=1in]{geometry}
\usepackage{amsmath}
\usepackage{upgreek}
\usepackage{xcolor}
\usepackage{graphicx}
\usepackage{upgreek}

\usepackage[utf8]{inputenc}

\usepackage{cite}

\usepackage{nameref,hyperref}

\usepackage{microtype}
\DisableLigatures[f]{encoding = *, family = * }

\raggedright
\usepackage{changepage}

\usepackage[aboveskip=1pt,labelfont=bf,labelsep=period,singlelinecheck=off]{caption}

\makeatletter
\renewcommand{\@biblabel}[1]{\quad#1.}
\makeatother

\usepackage{lastpage,fancyhdr,graphicx}
\usepackage{epstopdf}
\fancyheadoffset[L]{2.25in}
\fancyfootoffset[L]{2.25in}

\usepackage{color}

\definecolor{Gray}{gray}{.25}

\usepackage{graphicx}

\usepackage{sidecap}

\usepackage{wrapfig,cite}
\usepackage[pscoord]{eso-pic}
\usepackage[fulladjust]{marginnote}
\reversemarginpar

\usepackage{ragged2e}
\justifying\let\raggedright\justifying

\begin{document}

	\vspace*{0.35in}
	
	\begin{flushleft}
		{\Large
			\textbf\newline{Realization of broadband index-near-zero modes in nonreciprocal magneto-optical heterostructures}
		}
		\newline
		
		Yun Zhou,\textsuperscript{1,2,7} Panpan He,\textsuperscript{3,7} Sanshui Xiao,\textsuperscript{4} Fengwen Kang,\textsuperscript{4,5} Lujun Hong,\textsuperscript{6} Yun Shen,\textsuperscript{6} Yamei Luo,\textsuperscript{1,2}  and Jie Xu,\textsuperscript{1,2,*}
		
		\bigskip
		\bf{1} School of Medical Information and Engineering, Southwest Medical University, Luzhou 646000, China\\
		\bf{2} Medicine \& Engineering \& Information Fusion and Transformation Key Laboratory of Luzhou City, Luzhou 646000, China\\
		\bf{3} School of Electrical and Electronic Engineering, Luzhou Vocational \& Technical College, Luzhou 646000, China\\
		\bf{4} DTU Fotonik, Department of Photonics Engineering, Technical University of Denmark, DK-2800 Kgs. Lyngby, Denmark\\
		\bf{5} Laboratory of Advanced Nano Materials and Devices, Ningbo Institute of Materials Technology and Engineering (NIMTE), Chinese Academy of Sciences (CAS), Ningbo, 315201, China\\
		\bf{6} Institute of Space Science and Technology, Nanchang University, Nanchang 330031, China\\
		\bf{7} These authors contributed equally to this work\\
		\bf{*} xujie011451@163.com\\


	\end{flushleft}
	
	\section*{Abstract}
	Epsilon-near-zero (ENZ) metamaterial with the relative permittivity approaching zero has been a hot research subject in the past decades. The wave in the ENZ region has infinite phase velocity ($v=1/\sqrt{\varepsilon\mu}$), whereas it cannot efficiently travel into the other devices or air due to the impedance mismatch or near-zero group velocity. In this paper, we demonstrate that the tunable index-near-zero (INZ) modes with vanishing wavenumbers ($k=0$) and nonzero group velocities ($v_\mathrm{g} \neq 0$) can be achieved in nonreciprocal magneto-optical systems. This kind of INZ modes has been experimentally demonstrated in the photonic crystals at Dirac point frequencies and that impedance-matching effect has been observed as well. Our theoretical analysis reveals that the INZ modes exhibit tunability when changing the parameter of the one-way (nonreciprocal) waveguides. Moreover, owing to the zero-phase-shift characteristic and decreasing $v_\mathrm{g}$ of the INZ modes, several perfect optical buffers (POBs) are proposed in the microwave and terahertz regimes. The theoretical results are further verified by the numerical simulations performed by the finite element method. Our findings may open the new avenues for research in the areas of ultra -strong or -fast nonlinearity, perfect cloaking, high-resolution holographic imaging and wireless communications.

	
	\section{Introduction}
	Symmetry is common in natural world. For example, the roads of the incident and reflected light or sound are always symmetric. However, such a common characteristic cannot hold all the time in quantum physics. The chiral edge modes observed in quantum Hall effect, for instance, exhibit nonreciprocal propagation properties in opposite directions\cite{Prang:Th}. Similarly, the one-way electromagnetic (EM) modes are permitted to propagate unidirectionally, and they have been investigated and observed in many physical systems such as the magneto-optical (MO) heterostructure\cite{Tsakmakidis:Br,Shen:On,Wang:To}. The one-way EM modes in the MO-based structures refer as unidirectional (one-way) surface magnetoplasmons (SMPs) that are sustained at the surfaces of the MO materials. In 2009, the first experimental realization of unidirectional SMPs was achieved in a photonic crystal (PhC) consisting of yttriun-iron-garnet (YIG) rods\cite{Wang:Ob}. Recently, many striking works concerning the unidirectional SMPs have been reported. Tsakmakidis's group demonstrated that in the MO heterostructure the time-bandwidth limit, which was believed to be a fundamental limit in both engineering and theory, can be overcome due to the broken Lorentz reciprocity\cite{Tsakmakidis:Br}. More recently, our group reported several MO-based one-way nonreciprocal waveguides where many interesting phenomena such as slow wave\cite{Xu:Br}, truly trapping of robust unidirectional rainbow\cite{Xu:Sl,Xu:Re} and bidirectionally slowed down rainbow trapping and releasing\cite{Xu:Tr} were illustrated. It is inspiring to use the simple powerful MO heterostructure to demonstrate novel devices with new functionalities. 
	
	Metamaterials (MMs) with fantastic electromagnetic characteristics have been an attractive topic in the past decades. Many different MMs such as negative-index MMs\cite{Smith:Me,Yang:De}, hyperbolic metamaterials\cite{Huo:Hy,Li:Fu} and epsilon-negative MMs\cite{He:Ep} were proposed by several research groups. Besides, the index-near-zero (INZ) MMs, including the epsilon-near-zero (ENZ) MMs with vanishing permittivity\cite{Silveirinha:Tu,Zhou:Br,Davoyan:Op,Torres:Te}, the $\mu$-near-zero MMs with vanishing permeability\cite{Marcos:Ne}, double-near-zero (DNZ) MMs with both permittivity and permeability approach zero\cite{Ziolkowski:Pr,Nguyen:To,Xu:No}, and the INZ MMs proposed in the PhCs\cite{Huang:Di,Huang:Su}, are one of the research focus. There are three typical ways to build the ENZ MMs: 1) utilizing the plasma materials such as indium-tin-oxide (ITO) where the ENZ region is always near the plasma frequency\cite{Engheta:Pu,Alam:La}; 2) building the rectangle waveguide filled with magneto-optical materials while the ENZ region near the cutoff frequency\cite{Yi:Ep}; 3) fabricating layered metal-dielectric configurations and according to the effective medium theories (EMTs), the real part of the relative permittivity maybe vanish at discrete frequencies\cite{Suresh:En}. We note that the group velocities ($v_\mathrm{g}$) in most of the ENZ MMs tend to zero since the imaginary part of the permittivity is negligible as well, leading to forbidden energy transmission\cite{Javani:Re}. The double-near-zero materials (DNZ) with both the permittivity and permeability were suggested to solve such problem\cite{Alu:Ep,Ziolkowski:Pr,Lib:Ne}. However, both the ENZ and DNZ materials always suffer from the narrow operating band and complex manufacturing processes\cite{Kinsey:Ne}. On the other hand, the INZ materials have wide applications in energy harvesting based on supercoupling effect\cite{Engheta:Pu,Silveirinha:Th}, wavefront shaping\cite{Ziolkowski:Pr}, terahertz lens\cite{Torres:Te}, acoustic MMs or metasurface\cite{Jing:Nu,Li:Un,Zheng:Ac,Shen:As}, quantum optics\cite{Liberal:Ze}, switch of total transport and total reflection\cite{Liberal:Ne,Nguyen:To}. We emphasise that even though the experimental realization of the INZ modes has been proposed in all-dielectric system at Dirac point frequencies\cite{Moitra:Re}, it is still of significant interests digging novel broadband INZ materials with nonzero $v_\mathrm{g}$. 
	
	To this end, in what follows, we propose the broadband INZ modes with vanishing wavenumber ($k=0$) in the simple MO heterostructures. The explored INZ modes are similar to the Dirac point modes illustrated in the PhCs since the effective refractive indices of both modes are approaching zero. Moreover, based on our theoretical analysis, the working frequencies of such MO INZ modes can be continually adjusted by either modifying the thickness parameters of the structure or the external magnetic field. Therefore, compared to the previously reported INZ modes, the proposed broadband and adjustable INZ modes in this work are more favorable in many fields such as strong nonlinearity, cloaking, wireless communication and wavefront shaping. As a proof of concept, we further design several microwave and terahertz perfect optical buffers where the EM wave can be slowed down with zero phase shift. All the theoretical results are confirmed by numerical simulations using the finite element method.

	\section{Slowed down unidirectional SMPs}
	In our previous works\cite{Xu:Br,Xu:Sl,Xu:Ul,Xu:Tr}, cutting off the magnetic-optical (MO) material has been proved to be an efficient way to engineer the dispersion of the EM modes (e.g., SMPs) in the nonreciprocal systems. In this paper, we will show how such technique can be used to achieve broadband index-near-zero modes. We first study the propagation properties in a metal-dielectric-YIG-metal structure shown in Fig. 1(a). Since the metal can be treated as perfect electric conductor at microwave frequencies, such configuration is named the 'EDYE'. Note that, the dielectric in this subsection is set to be air with relative permittivity $\varepsilon_r=1$. An external magnetic field ($H_0$) was applied on the YIG layer in $-z$ direction and the permeability of YIG in this case has the form\cite{Pozar:Mi}
	\begin{equation}
		\stackrel{\rightarrow}{\mu}=\left[\begin{array}{ccc}
			\mu_{1} & -i \mu_{2} & 0 \\
			i \mu_{2} & \mu_{1} & 0 \\
			0 & 0 & 1
		\end{array}\right],
	\end{equation}
	\begin{equation}
		\mu_{1}=1+\frac{\omega_\mathrm{m}\left(\omega_{0}-i \nu \omega \right)}{\left(\omega_{0}-i \nu \omega\right)^{2}-\omega^{2}},	
		\mu_{2}=\frac{\omega_\mathrm{m} \omega}{\left(\omega_{0}-i \nu \omega\right)^{2}-\omega^{2}}.
	\end{equation}
	where $\omega$, $\nu$, $\omega_0$ and $\omega_\mathrm{m}$ are the angular frequency, the damping coefficient, the precession angular frequency and the characteristic circular frequency, respectively. $\omega_0=2\pi\gamma H_0$ and $\gamma=2.8\times10^6$ rad/s/G is the gyromagnetic ratio. According to Maxwells' equations and the boundary conditions of the EDYE configuration, the dispersion relation of the SMPs can be described by the following equation
	\begin{figure}[hpt]
		\centering\includegraphics[width=4.5 in]{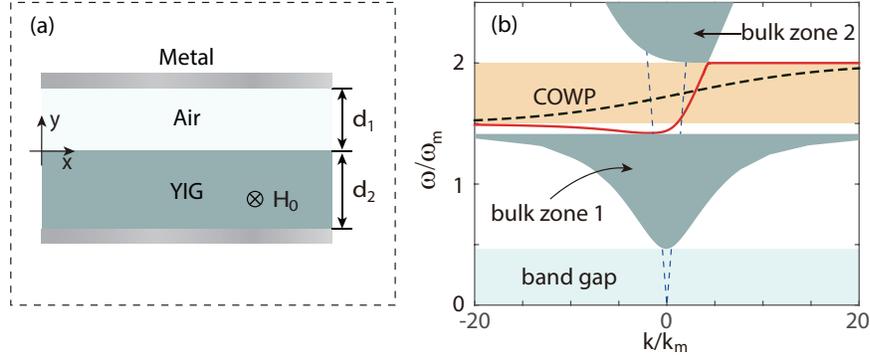}
		\caption{ (a) The schematic of the EDYE model. (b) The dispersion diagrams of SMPs in the metal-dielectric-YIG-metal (EDYE) structure as $D=D_1=(0.1\lambda_\mathrm{m},0.15\lambda_\mathrm{m})$ (red line) and $D=D_2=(0.01\lambda_\mathrm{m},0.01\lambda_\mathrm{m})$ (dashed black line). The yellow shaded areas indicate the complete one-way propagation (COWP) band while the gray and cyan shaded regions respectively demonstrate the compressed bulk zones and band gaps in the $D_1$ case. The blue dashed lines are light lines. The other parameters are $\varepsilon_r=1$, $\omega_0=\omega_\mathrm{m}$, $\varepsilon_\mathrm{m}=15$ and $\omega_\mathrm{m}=2\pi\times 5\times 10^9$ rad/s. }\label{Fig1}
	\end{figure}
	
	\begin{equation}
		\alpha_\mathrm{d} \mu_\mathrm{v}+\alpha_\mathrm{v}\frac{\tanh\left(\alpha_\mathrm{d} d_1\right)}{\tanh \left(\alpha_\mathrm{v} d_2\right)} +\frac{\mu_{2}}{\mu_{1}} k \tanh \left(\alpha_\mathrm{d} d_1\right)=0
	\end{equation}
	where $d_1$ and $d_2$ are the thicknesses of the dielectric and the YIG layer. For simplicity, we introduce a thickness parameter $D=(d_1,d_2)$ in this paper. $\alpha_\mathrm{d}=\sqrt{k^2-\varepsilon_r k_0^2}$ and $\alpha_\mathrm{v}=\sqrt{k^2-\varepsilon_\mathrm{m} \mu_\mathrm{v} k_0^2}$ are the attenuation coefficients of SMPs in the dielectric and the YIG layer. We emphasise here that the YIG parameters used in this paper are $\varepsilon_\mathrm{m}=15$ (the permittivity of YIG) and $\omega_\mathrm{m}=2\pi\times5\times 10^9$ rad/s ($f_\mathrm{m}=5$ GHz). Owing the third item of Eq. (3), the propagation characteristic of the EM modes with $k>0$ and $k<0$ are different. As it is well established, the COWP bands in the one-way waveguides are heavily dependent on the asymptotic frequencies (AFs). According to Eq. (3), we can easily calculate the AFs and they can be written as following
	\begin{equation}
		\omega_\mathrm{AF}^-=\omega_0+\frac{\omega_\mathrm{m}}{2}, \quad \text { ($k \to -\infty$) }
	\end{equation}
	\begin{equation}
		\omega_\mathrm{AF}^+=\omega_0+\omega_\mathrm{m}. \quad \text { ($k \to +\infty$)}
	\end{equation}

	In Fig.1 (b), two different thickness parameters (D) were considered to show the dispersion properties in the EDSE configurations. The red line indicates the dispersion curve of the SMPs as $D=D_1=(0.1\lambda_\mathrm{m},0.15\lambda_\mathrm{m})$ ($\lambda_\mathrm{m}=c/f_\mathrm{m}\approx 60$ mm, c is the speed of light in vacuum), and $\omega_0=\omega_\mathrm{m}$ ($H_0\approx 1785.7$G). It is clear that there is a complete one-way propagation (COWP) band (the yellow shaded area) which is limited by two AFs, i.e. $\omega_\mathrm{AF}^-$ and $\omega_\mathrm{AF}^+$. The gray colored areas represent the bulk zones of YIG, which are much narrower than the ones in the infinite YIG cases\cite{Deng:On}, and a new band gap (the cyan zone) emerged below the compressed bulk zone 1. As a contrast, the dashed black line demonstrates the dispersion diagram in a ultra-thin EDYE structure with $D=D_2=(0.01\lambda_\mathrm{m},0.01\lambda_\mathrm{m})$. In this case, the bandwidth of the COWP band is analogous to previous one in the $D_1$ case since the AFs remain the same. We note that the bulk zones 1 and 2 are both narrowed down (not shown in Fig. 1(b)) due to the ultra-thin vertical width\cite{Xu:Br}, and broad band gap region were observed as well, which should has potential uses in design of optical isolator. The dispersion curve of the SMPs (red line) is much more horizontal than the one in Fig. 1(a), implying smaller group velocities and slowed down EM waves. 
	\begin{figure}[tb]
		\centering\includegraphics[width=4.5 in]{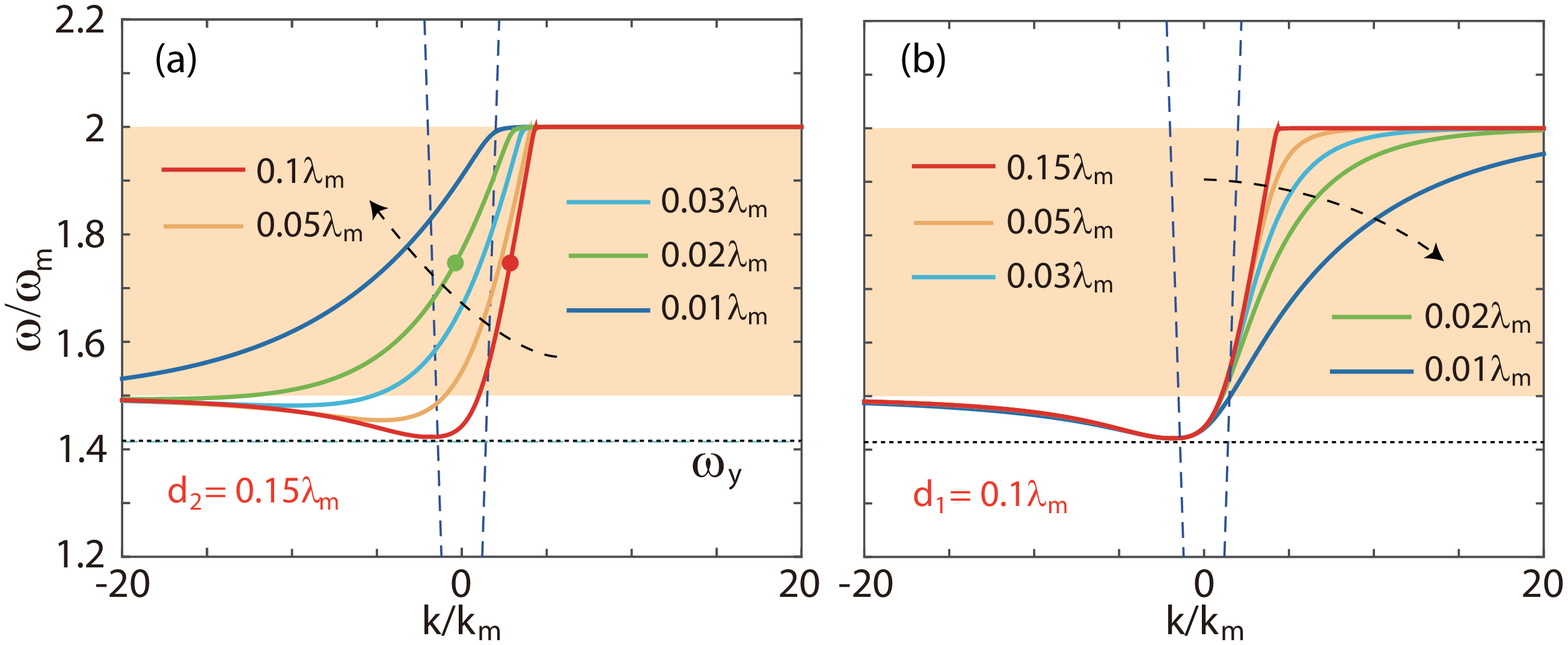}
		\caption{ The relation between the thickness parameter $D=(d_1,d_2)$ and the dispersion curves of SMPs. (a) The dispersion diagrams for five different values of $d_1$ when $d_2=0.15\lambda_\mathrm{m}$. (b) The dispersion diagrams for five different values of $d_2$ when $d_1=0.1\lambda_\mathrm{m}$.}\label{Fig2}
	\end{figure}	
	\begin{figure}[ptb]
		\centering\includegraphics[width=5 in]{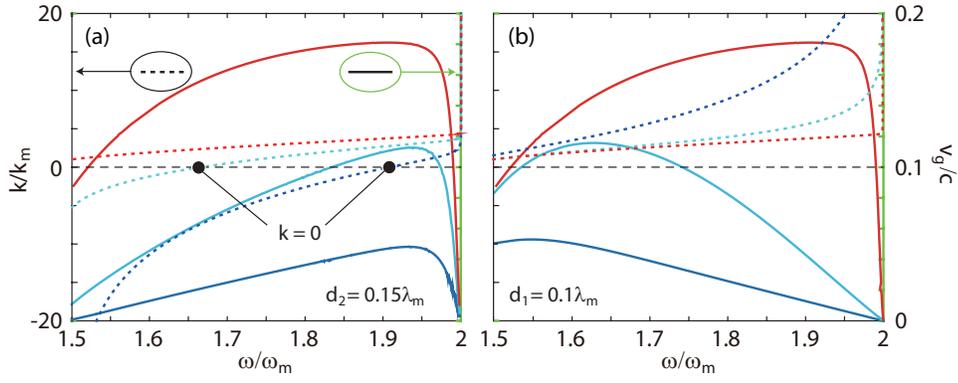}
		\caption{ The wavenumbers (k) and the group velocities ($v_\mathrm{g}$) of SMPs (a) in Fig.  2(a) and (b) in Fig. 2(b) as functions of $\omega$. The solid lines represent the values of $v_\mathrm{g}$ whereas the dotted lines shows the corresponding values of k. The other parameters are the same as in Fig. 1.}\label{Fig3}
	\end{figure}
	
	To clarify the slow-wave phenomenon occurred in the ultra-thin EDYE structure, we investigated two cases of the EDYE structure. In the first case, constant $d_2$ ($=0.15\lambda_\mathrm{m}$) and five different $d_1$ are considered. The corresponding dispersion diagrams around the COWP region are demonstrated in Fig. 2(a), in which we found that when reducing the thickness of the air layer ($d_1$), the dispersion branches with $k<0$ rose up distinctly while most of the dispersion branches in $k>0$ region located near $\omega=\omega_\mathrm{AF}^+$. In the second case, we considered the constant $d_1$ ($=0.1\lambda_\mathrm{m}$) and five different values of $d_2$, i.e. $0.15\lambda_\mathrm{m}$, $0.05\lambda_\mathrm{m}$, $0.03\lambda_\mathrm{m}$, $0.02\lambda_\mathrm{m}$ and $0.1\lambda_\mathrm{m}$. We plotted the corresponding dispersion diagram as shown in Fig. 2(b). Different from the dispersion curves shown in Fig. 2(a), when decreasing the value of $d_2$, the dispersion branches with $k>0$ descend down whereas the dispersion branches with $k<0$ are nearly the same. It is obvious that the group velocities of SMPs decreased in both cases when decreasing the value of $d_1$ (see Fig. 2(a)) or $d_2$ (see Fig. 2(b)). More clearly results are displayed in Fig. 3, in which the corresponding group velocities ($v_\mathrm{g}$) and wavenumber (k) of the SMPs are demonstrated in the COWP band $1.5\omega_\mathrm{m}<\omega<2\omega_\mathrm{m}$. In Fig. 3(a), the red, the cyan and the blue solid lines respectively represent the values of $v_\mathrm{g}$ in the cases of $d_1=0.1\lambda_\mathrm{m}$, $d_1=0.03\lambda_\mathrm{m}$ and $d_1=0.01\lambda_\mathrm{m}$ when $d_2=0.15\lambda_\mathrm{m}$. As expected, when decreasing the value of $d_1$, the solid lines dropped down. Moreover, similar results are shown in Fig. 3(b) for constant $d_1$ cases. On the other hand, it is quite different between constant $d_2$ cases and constant $d_1$ cases in values of k of the unidirectional modes. As shown in Fig. 3, the dashed lines illustrate the changing of k of the SMPs. In the case of $d_1=0.1\lambda_\mathrm{m}$, the wavenumbers of SMPs in the COWP band are constantly greater than zero. On the contrary, for $d_2=0.15\lambda_\mathrm{m}$ case, the wavenumbers can be positive or negative depending on the values of $d_2$ and frequency. More excitingly, as demonstrated in Fig. 3(a), the EM modes (SMPs) with $k=0$ (marked by points) and $v_\mathrm{g}\neq0$, i.e. the index-near-zero (INZ) modes, were found in the EDYE structures. Deriving from Figs. 2 and 3, it is believable that the INZ modes can be manipulated in the whole COWP band by engineering the thickness of air ($d_1$). In another word, the EDYE configuration, as discussed above, can support the broadband INZ modes by simply engineering the thickness parameter of the structures. Therefore,  due to the slow-wave and broadband characteristics, such simple configurations are promising for designing optical functional devices, and one of its directly uses is designing the perfect optical buffer (POB) with near-zero phase shift. It is also worth to note that as presented in our previous works, varying the external magnetic field should be another efficient way to engineering the INZ modes\cite{Xu:Re,Xu:Br}.

	\begin{figure}[htb]
		\centering\includegraphics[width=4.5 in]{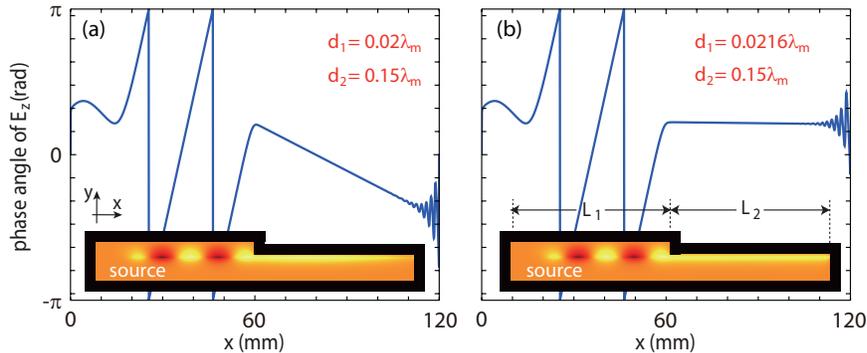}
		\caption{ The phase angle ($\Phi$) of $E_z$ along the air-YIG interface in the finite element method (FEM) simulations for (a) $D=D_3=(0.02\lambda_\mathrm{m},0.15\lambda_\mathrm{m})$, and (b) $D=D_4=(0.0216\lambda_\mathrm{m},0.15\lambda_\mathrm{m})$. The insets show the distributions of the simulated electric field ($E_z$). The other parameters are $f=1.75f_\mathrm{m}$, $L_1=L_2=60$ mm and $\nu=10^{-3}\omega$.}\label{Fig4}
	\end{figure}
	\section{Index-near-zero modes and perfect optical buffer}
	Zero phase shift (ZPS) is an important characteristic in many fields, for example, the near-field radio frequency identification (RFID) applications\cite{Shi:El,Zeng:Di}. In this subsection, we will propose the two-dimensional (2D) and three-dimensional (3D) POBs in the microwave regime based on the EDYE configuration. The inset of Fig. 4(a) indicates the simulated magnetic field ($E_z$) in a thick-thin straight EDYE waveguide as $f=1.75f_\mathrm{m}$ and $\nu=10^{-3}\omega$ (the damping coefficient). The thickness parameters for the thick and thin parts are respectively $D=D_1=(0.1\lambda_\mathrm{m},0.15\lambda_\mathrm{m})$ and $D=D_3=(0.02\lambda_\mathrm{m},0.15\lambda_\mathrm{m})$, and two parts have the same length $L_1=L_2=60$ mm. We note here that the simulations were performed by utilizing finite element method (FEM) and the lossy MO materials were considered in all simulations to illustrate the loss effects on the nonreciprocal and the INZ characteristics. In the simulations, the EM mode was excited by a line current source at $x=20$ mm in the thick part and it propagated along the air-YIG interface with the phase angle ($\Phi$) of $E_z$ changing in the $[-\pi,\pi]$ region. Once the wave reaching and further transmitting into the thin EDYE part, the curve (blue solid line in Fig. 4(a)) of $\Phi$ of $E_z$ becomes gentle since $|k|$ of the SMPs in the thick part is larger than the one in the thin part (see points in Fig. 2). It is quite inspiring if we can manipulate the phase-angle curve and make it to be horizontal, which implies ZPS. In Fig. 4(b), we slightly enlarged $d_1$ of the thin part and set $D=D_4=(0.0216\lambda_\mathrm{m},0.15\lambda_\mathrm{m})$. Excitingly, $\Phi$ remains almost unchanged in the thin part except for the area near the rightmost end (PEC wall) of the waveguide. Since the EM wave was slowed down in the thin part and the clear ZPS has been observed in the slow-wave thin part, the structure demonstrated in the inset of Fig. 4(b) is suitable for design of the POB. We note here that in our previous work, we have shown that the surface wave located in the light cone could efficiently couple with the fundamental modes of traditional dielectric waveguide\cite{Shen:On}. Therefore, the above mentioned INZ modes are promising in many fields such as wireless communication and optical cloaking. 
	
	\begin{figure}[tb]
		\centering\includegraphics[width=5 in]{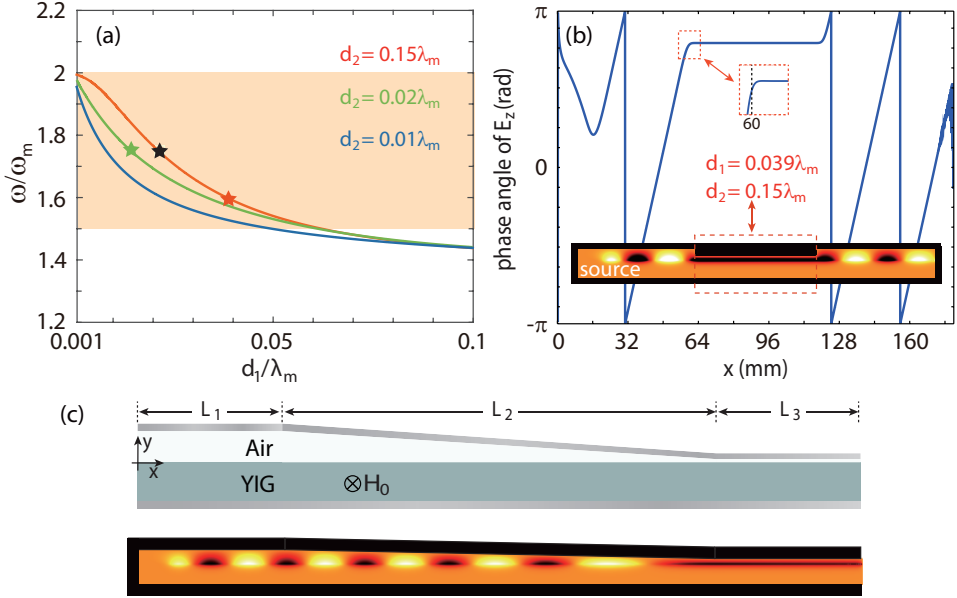}
		\caption{ (a) Theory of the broadband perfect optical buffer (POB) based on EDYE configuration. Solid lines illustrate the numerical solutions of $k=0$ when $d_2=0.15\lambda_\mathrm{m}$ (red line), $d_2=0.02\lambda_\mathrm{m}$ (green line) and $d_2=0.01\lambda_\mathrm{m}$ (blue line), respectively. (b) The phase angle of $E_z$ along the air-YIG interface in a microwave POB for $f=1.6f_\mathrm{m}$. The thickness parameter in the compressing part is $D=D_5=(0.039\lambda_\mathrm{m},0.15\lambda_\mathrm{m})$ (red star in (a)). The inset illustrates the distribution of simulated $E_z$. (c) The schematic and the FEM simulation of a tapered EDYE structure.}\label{Fig5}
	\end{figure}
	
	To further explore the POB theory based on the EDYE configuration, the frequencies ($\omega$) of the INZ modes are plotted as a function of $d_1$ for three different $d_2$, and the results are demonstrated in Fig. 5(a). As one can see, the black star is on the red line which indicates the case of $d_2=0.15\lambda_\mathrm{m}$, and the corresponding frequency $\omega=1.75\omega_\mathrm{m}$ while $d_1\approx0.0216\lambda_\mathrm{m}$. Thus, our calculation fit well with the simulation results illustrated in Fig. 4(b). The green and blue lines indicate the cases of smaller $d_2$, i.e. $d_2=0.02\lambda_\mathrm{m}$ and $d_2=0.01\lambda_\mathrm{m}$. It is obvious that when decreasing $d_2$, the line dropped down which implies that the calculated $\omega$ of the INZ modes decreased. One can derive the same conclusion from Fig. 2(b), in which the dispersion curve dropped down when decreasing $d_2$. Noteworthily, the INZ region can be efficiently varied in the COWP band by modifying the thickness parameter while the COWP band can be tuned by changing the external magnetic field\cite{Shen:Tr}. The INZ modes/region mentioned here can be dynamically switched from 'on' to 'off' in practice by controlling the external magnetic field. In this work, we just focus on the cases of stable external magnetic field. Based on the above results, we further designed a complete POB consisting of three straight EDYE parts. As shown in the inset of Fig. 5(b), the values of $d_1$ in the three parts are $d_1=0.1\lambda_\mathrm{m}$, $d_1=0.039\lambda_\mathrm{m}$ and $d_1=0.1\lambda_\mathrm{m}$, respectively. In the simulation, we set the operating frequency $f=1.6f_\mathrm{m}$ (red star in Fig. 5(a)). The launched wave, as deposited in Fig. 5(b), transmitted through the second part and recovered in the third part with the phase angle nearly unaffected by the propagation process in the second part. Due to the coupling effect between the modes in the thick and thin parts, slight phase shift was found at the thick-thin interfaces (see the upper inset of Fig. 5(b)). A efficient way to reduce the coupling effect is adding a tapered EDYE configuration between the thick and thin EDYE parts. As shown in Fig. 5(c), we designed a thick-tapered-thin (TTT) EDYE structure. In the simulations of the TTT structure, we found that the effective wavelength ($\lambda_\mathrm{ef}$) gradually lengthened along the $+x$ direction in the tapered part and $\lambda_\mathrm{ef}$ tend to near infinite in the ultra-thin part. More importantly, the phase shift induced by the propagation in the ultra-thin part can be less than $0.3\% \Phi_0$ ($\Phi_0$ is the incident phase angle) when $L_2=3L_1=180$ $\upmu$m. 
	
	\begin{figure}[t]
		\centering\includegraphics[width=5 in]{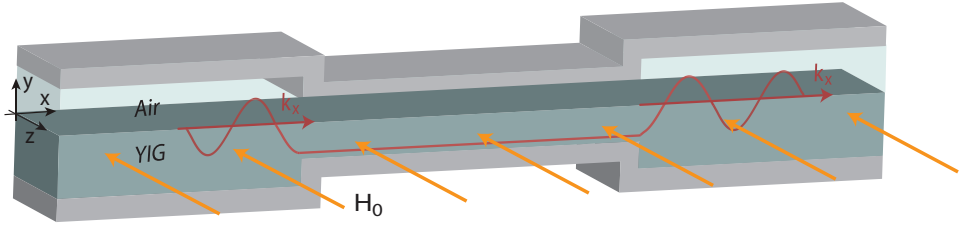}
		\caption{ Schematic of a three-dimensional (3D) POB.}\label{Fig6}
	\end{figure}
	\begin{figure}[t]
		\centering\includegraphics[width=5 in]{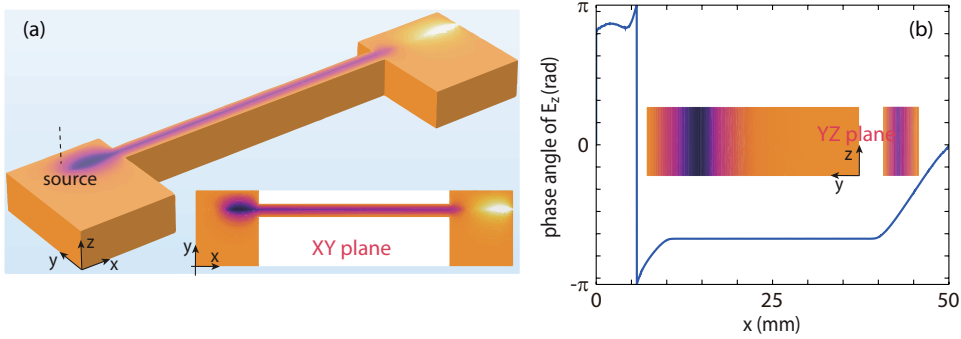}
		\caption{ (a) Simulated electric-field distribution in a 3D POB where the thickness parameters of the thick and thin parts are respectively $D=D_6=(0.05\lambda_\mathrm{m},0.15\lambda_\mathrm{m})$ and $D=D_7=(0.01452\lambda_\mathrm{m},0.02\lambda_\mathrm{m})$. The inset illustrates the distribution of $E_z$ in XY plane ($z=\mathrm{w}/2$) and the inset of (b) shows the corresponding distribution of $E_z$ in YZ planes (Left: $x=7$ mm. Right: $x=17$ mm.). (b) The phase angle of $E_z$ along the interface of the XY plane ($z=\mathrm{w}/2$) and the XZ plane ($y=d_{10}/2$). }\label{Fig7}
	\end{figure}
	The 2D POB theory based on the EDYE configuration can also be used to design realistic 3D POBs. The schematic of a designed 3D POB is shown in Fig. 6, which is also constructed by three EDYE parts and the direction of the external magnetic field is assumed to be $-z$. According to the analysises of Figs. 2 and 3, different from the above 2D POB, both the air and YIG layers of the thin part of the 3D POB are designed to be narrower than the thick parts to efficiently slow the wave. As an example, the thickness parameters in the three parts of the 3D POB are respectively set to be $D=D_6=(0.05\lambda_\mathrm{m},0.15\lambda_\mathrm{m})$, $D=D_7=(0.01452\lambda_\mathrm{m},0.02\lambda_\mathrm{m})$ (marked by the green star in Fig. 5(a)) and $D=D_6$, and we further performed the FEM simulation in such 3D EDYE structure. As shown in Fig. 7(a) and the inset of Fig. 7(b), similar to the 2D simulations presented in Figs. 4 and 5, the wave unidirectionally propagated from the first thick part and traveled through the ultra-thin part and was trapped in the rightmost end of the last part. Fig. 7(b) demonstrates the changing of the phase angle of the wave along $y=d_{10}/2$ ($d_{10}$, the air thickness in the thin part) in the XY plane ($z=\mathrm{w}/2$ where w is the lateral thickness of the structure). As expected, clear ZPS was also observed in the thin part of such 3D one-way waveguide. In a sence, our designed microwave POB can work as a type of cloaking device since the EM wave travels through the structure with ZPS and without reflection.

	\section{The INZ mode-based POB in the terahertz regime}
	Nowadays, communication band gradually transfers from microwave to terahertz regime due to the increase in demand for enormous information. It is really meaningful if the POB theory revealed above can be applied in the terahertz regime. As we reported in previous works\cite{Xu:Br,Xu:Tr,Xu:Sl}, the COWP band may be narrowed down or even vanish in sub-wavelength one-way terahertz waveguides. Moreover, to our knowledge, the INZ modes have never been reported in the one-way terahertz waveguide sandwiched by metals. In this subsection, we propose two types of terahertz POB utilizing perfect magnetic conductor (PMC) boundary. As shown in Fig. 8, the leftmost two diagrams indicate the two kinds of one-way terahertz waveguides and the upper one is the PMC-dielectric-semiconductor-PEC (MDSE) structure while the lower one is the PMC-dielectric-semiconductor-PMC (MDSM) structure. Here, we note that the dielectric layers applied in the MDSE and MDSM waveguides are respectively air and silicon for engineering  appropriate COWP band. We emphasize that the PEC-dielectric-semiconductor-PEC (EDSE) and PEC-dielectric-semiconductor-PMC (EDSM) structures, according to our analysis, cannot sustain the INZ modes, thus they were not considered in designing terahertz POB. The dispersion relation of SMPs in the MDSE and MDSM structures can be derived by solving the Maxwell's equations and they have the following forms
	\begin{equation}	
		\left(k^{2}-\varepsilon_{1} k_{0}^{2}\right) \tanh \left(\alpha d_{2}\right)+\frac{\varepsilon_{1}}{\varepsilon_{\mathrm{d}}} \alpha_{\mathrm{d}}\left[\alpha-\frac{\varepsilon_{2}}{\varepsilon_{1}} k \tanh \left(\alpha d_{2}\right)\right] \frac{1}{\tanh \left(\alpha_{\mathrm{d}} d_{1}\right)}=0, \quad \text { (MDSE) }
	\end{equation}
	\begin{equation}
		\frac{\varepsilon_{2}}{\varepsilon_{1}} k+\frac{\alpha}{\tanh \left(\alpha d_{2}\right)}+\frac{\varepsilon_\mathrm{v}}{\varepsilon_{\mathrm{d}}} \frac{\alpha}{\tanh \left(\alpha_{\mathrm{d}} d_{1}\right)}=0. \quad \text { (MDSM) }		
	\end{equation}
	where $\varepsilon_{1}=\varepsilon_{\infty}\left\{1-\frac{(\omega+i v) \omega_\mathrm{p}^{2}}{\omega\left[(\omega+i v)^{2}-\omega_\mathrm{c}^{2}\right]}\right\}$, $\varepsilon_{2}=\varepsilon_{\infty} \frac{\omega_\mathrm{c} \omega_\mathrm{p}^{2}}{\omega\left[(\omega+i v)^{2}-\omega_\mathrm{c}^{2}\right]}$ and $\varepsilon_\mathrm{v}=\varepsilon_1-\frac{\varepsilon_2^2}{\varepsilon_1}$. $\varepsilon_\infty$, $\nu$, $\omega_\mathrm{p}$ and $\omega_\mathrm{c}=\frac{eH_0}{m^*}$ are the high-frequency permittivity, the electron scattering frequency, the plasma frequency and the electron cyclotron frequency of the semiconductor. $\varepsilon_\mathrm{d}$, $\alpha$ and $\alpha_\mathrm{d}$ are the relative permittivity of the dielectric, the attenuation coefficient in the semiconductor and the attenuation coefficient in the dielectric, respectively. The semiconductor in this paper is assumed to be N-type InSb with $\omega_\mathrm{p}=4 \pi \times 10^{12}$ rad/s and $\varepsilon_\infty=15.6$. The asymptotic frequency (AF) is one of the main key in the study of one-way waveguide and by solving Eqs. 6 and 7, we found that there are three AFs in the MDSE model and four AFs in the MDSM model. The AFs in the MDSE structure can be written as
	\begin{equation}
		\begin{cases}
			\omega_{\mathrm{AF}}^{(1)}=\omega_\mathrm{c} \\
			\omega_{\mathrm{AF}}^{(2)}=\omega_{\mathrm{sp}}^{(1)}=\frac{1}{2}\left(\sqrt{\omega_{\mathrm{c}}^{2}+\frac{4 \varepsilon_{\infty}}{\varepsilon_{\infty}+\varepsilon_{\mathrm{d1}}} \omega_{\mathrm{p}}^{2}}-\omega_{\mathrm{c}}\right) \\
			\omega_{\mathrm{AF}}^{(3)}=\omega_{\mathrm{sp}}^{(2)}=\frac{1}{2}\left(\sqrt{\omega_{\mathrm{c}}^{2}+\frac{4 \varepsilon_{\infty}}{\varepsilon_{\infty}+\varepsilon_{\mathrm{d1}}} \omega_{\mathrm{p}}^{2}}+\omega_{\mathrm{c}}\right)
		\end{cases}
	\end{equation}
	while the AFs in the MDSM structures are found to be
	\begin{equation}
		\begin{cases}
			\omega_{\mathrm{AF}}^{(1)}=\omega_{\mathrm{sp}}^{(1)}=\frac{1}{2}\left(\sqrt{\omega_{\mathrm{c}}^{2}+\frac{4 \varepsilon_{\infty}}{\varepsilon_{\infty}+\varepsilon_{\mathrm{d2}}} \omega_{\mathrm{p}}^{2}}-\omega_{\mathrm{c}}\right) \\
			\omega_{\mathrm{AF}}^{(2)}=\omega_{a}=\frac{1}{2}\left(\sqrt{\omega_{\mathrm{c}}^{2}+4 \omega_{\mathrm{p}}^{2}}-\omega_{\mathrm{c}}\right) \\
			\omega_{\mathrm{AF}}^{(3)}=\omega_{\mathrm{sp}}^{(2)}=\frac{1}{2}\left(\sqrt{\omega_{\mathrm{c}}^{2}+\frac{4 \varepsilon_{\infty}}{\varepsilon_{\infty}+\varepsilon_{\mathrm{d2}}} \omega_{\mathrm{p}}^{2}}+\omega_{\mathrm{c}}\right) \\	\omega_{\mathrm{AF}}^{(4)}=\omega_{b}=\frac{1}{2}\left(\sqrt{\omega_{\mathrm{c}}^{2}+4 \omega_{\mathrm{p}}^{2}}+\omega_{\mathrm{c}}\right) \\
		\end{cases}
	\end{equation}
	$\varepsilon_\mathrm{d1}=1$ and $\varepsilon_\mathrm{d2}=11.68$ are the relative permittivities of air and silicon, respectively. 
	\begin{figure}[t]
		\centering\includegraphics[width=5 in]{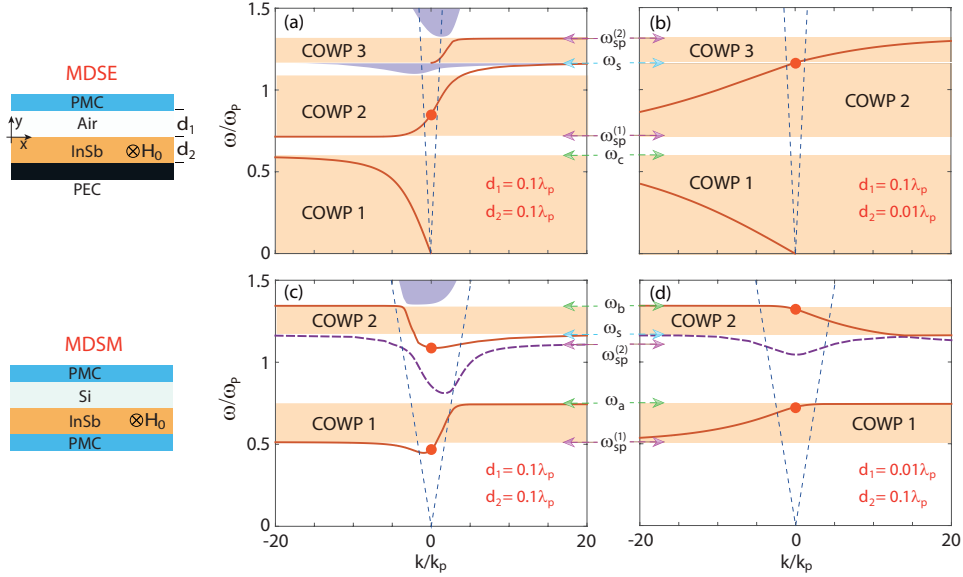}
		\caption{ Realization of the terahertz POB by utilizing PMC walls. The dispersion diagrams of PMC-air-InSb-PEC (MDSE) structures as (a) $D=D_8=(d_1,d_2)=(0.1\lambda_\mathrm{p},0.1\lambda_\mathrm{p})$ and (b) $D=D_9=(0.1\lambda_\mathrm{p},0.01\lambda_\mathrm{p})$. (c) and (d) demonstrate the dispersion diagrams of PMC-air-InSb-PMC (MDSM) structures as $D=D_{8}$ and $D=D_{10}=(0.01\lambda_\mathrm{p},0.1\lambda_\mathrm{p})$, respectively. Two leftmost diagrams show the schematics of the MDSE and MDSM configurations. The other parameters are $\varepsilon_\mathrm{d1}=1$, $\varepsilon_\mathrm{d2}=11.68$, $\varepsilon_\infty=15.6$ and $\omega_\mathrm{c}=0.6\omega_\mathrm{p}$. }\label{Fig8}
	\end{figure}
	
	Fig. 8(a) shows the dispersion curves of SMPs in the MDSE structure as $D=D_8=(0.1\lambda_\mathrm{p},0.1\lambda_\mathrm{p})$ ($\lambda_\mathrm{p}=150$ $\upmu$m) and $\omega_\mathrm{c}=0.6\omega_\mathrm{p}$. It is clear that there are three discrete COWP bands and three corresponding AFs are $\omega_{\mathrm{AF}}^{(1)}=0.6\omega_\mathrm{p}$, $\omega_{\mathrm{AF}}^{(2)} \approx0.71\omega_\mathrm{p}$ and $\omega_{\mathrm{AF}}^{(3)}\approx1.31\omega_\mathrm{p}$. The bulk zones are significantly compressed in this case, making a remarkable total bandwidth of the COWP bands. More excitingly, a terahertz INZ mode (marked by a red point) with $k=0$ was found in the COWP 2 region. When further decreasing the thickness of InSb, as shown in Fig. 8(b) where $D=D_9=(0.1\lambda_\mathrm{p},0.01\lambda_\mathrm{p})$, the bulk zones nearly disappear in the $[0,1.5\omega_\mathrm{p}]$ band. Moreover, the SMPs sustained at the air-InSb and PEC-InSb interfaces strongly coupled with each other when $\omega \approx \omega_\mathrm{s}$ ($\omega_\mathrm{s}=\sqrt{\omega_\mathrm{c}^2+\omega_\mathrm{p}^2}$), leading to the connecting dispersion curve (the upper red line). Besides, compared to the dispersion curves in the COWP 2 band in Fig. 8(a), the one in Fig. 8(b) rose up, resulting in the larger working frequency of the one-way INZ mode. We further plotted the dispersion diagrams of the MDSM structures for $D=D_{8}=(0.1\lambda_\mathrm{p},0.1\lambda_\mathrm{p})$ (Fig. 8(c)) and $D=D_{10}=(0.01\lambda_\mathrm{p},0.1\lambda_\mathrm{p})$ (Fig. 8(d)) as $\omega_\mathrm{c}=0.6\omega_\mathrm{p}$. Four AFs in this condition are respectively $\omega_{\mathrm{AF}}^{(1)} \approx0.51\omega_\mathrm{p}$, $\omega_{\mathrm{AF}}^{(2)} \approx0.74\omega_\mathrm{p}$, $\omega_{\mathrm{AF}}^{(3)} \approx1.11\omega_\mathrm{p}$ and $\omega_{\mathrm{AF}}^{(4)} \approx1.34\omega_\mathrm{p}$. As illustrated in Figs. 8(c) and 8(d), there are two COWP bands in which the EM waves have opposite propagation directions. More interestingly, the INZ modes can be found in all COWP bands once we carefully design the thickness of silicon (Fig. 8(d)). Since we have observed the INZ modes in the MDSE and MDSM waveguides, and the frequency of the INZ modes can be manipulated by controlling the thickness of semiconductor or dielectric, we believe that broadband tunable terahertz POB can be achieved based on such configurations. 
	\begin{figure}[t]
		\centering\includegraphics[width=6 in]{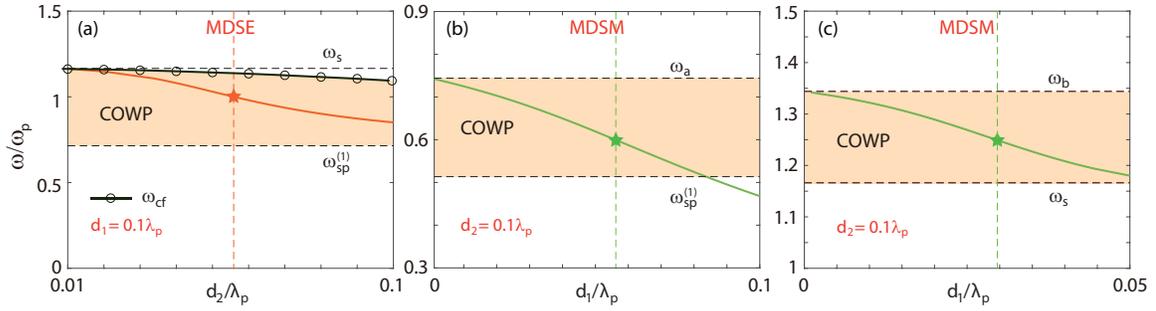}
		\caption{ (a) The frequencies of the INZ modes in the MDSE waveguides as a function of $d_2$ as $d_1=0.1\lambda_\mathrm{p}$. The frequencies of (b) the lower INZ modes and (c) the upper INZ modes in the MDSM waveguides as functions of $d_1$ as $d_2=0.1\lambda_\mathrm{p}$. The other parameters are the same as in Fig. 8.}\label{Fig9}
	\end{figure}
	\begin{figure}[t]
		\centering\includegraphics[width=6 in]{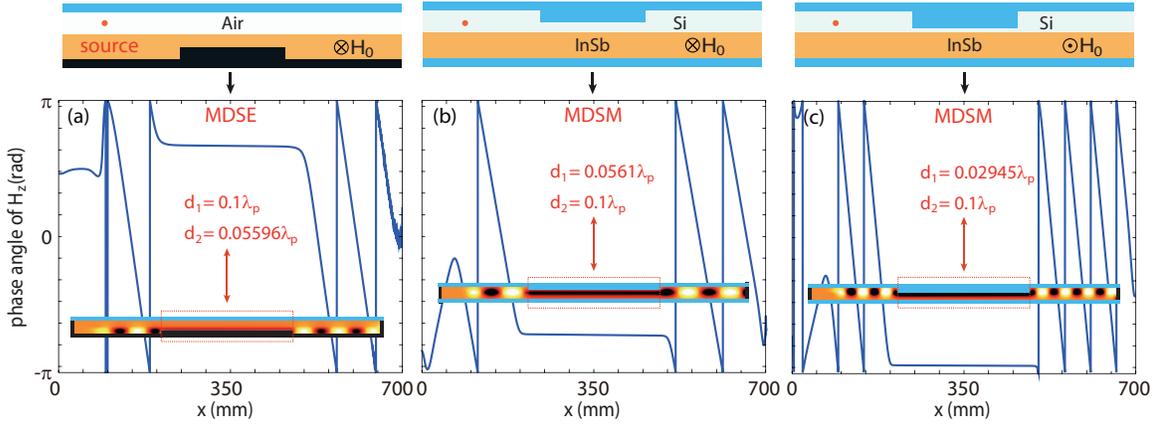}
		\caption{ The phase angle of $H_z$ along the dielectric-InSb interface in (a) MDSE-based and (b,c) MDSM-based terahertz POBs. The thickness parameters of the thin parts of the POBs are respectively (a) $D=D_{11}=(0.1\lambda_\mathrm{p},0.05596\lambda_\mathrm{p})$, (b) $D=D_{12}=(0.0561\lambda_\mathrm{p},0.1\lambda_\mathrm{p})$ and (c) $D=D_{13}=(0.02945\lambda_\mathrm{p},0.1\lambda_\mathrm{p})$. The operating frequencies in the simulations are $f=f_\mathrm{p}$, $f=0.6f_\mathrm{p}$ and $f=1.25f_\mathrm{p}$ in sequence. The InSb is considered to be lossy in the simulations with electron scattering frequency $\nu=0.001$.}\label{Fig10}
	\end{figure}
	
	In order to further demonstrate the relation between the INZ modes and the thickness parameters, we plotted $\omega$ of the INZ modes as a function of $d_2$ in Fig. 9(a) for the MDSE waveguide with $d_1=0.1\lambda_\mathrm{p}$. As we expected, the unidirectional INZ modes were found in a broad band within the COWP band. Note that the upper limit of the COWP band in Fig. 9(a) represent the cut-off frequency ($\omega_\mathrm{cf}$) of the lower bulk zone (see Fig. 8(a)). For the MDSM configuration, two kinds of INZ modes, i.e. the higher and the lower INZ modes (see Figs. 8(c,d)), were investigated. Consequently, the INZ modes were proved to be tunable in almost the entire COWP bands. Based on the above results, we designed three types of terahertz POBs which are named the type-1, type-2 and type-3 POBs, and the schematics of these POBs are exhibited in the top of Fig. 10. The thickness parameters of the thin parts of three POBs are respectively $D=D_{11}=(0.1\lambda_\mathrm{p},0.05596\lambda_\mathrm{p})$ (the vertical line in Fig. 9(a)), $D=D_{12}=(0.0561\lambda_\mathrm{p},0.1\lambda_\mathrm{p})$ (the vertical line in Fig. 9(b)) and $D=D_{13}=(0.02945\lambda_\mathrm{p},0.1\lambda_\mathrm{p})$ (the vertical line in Fig. 9(c)). The simulated magnetic field $H_z$ of three thick-thin-thick waveguides are shown in the insets of Fig. 10 and the working frequencies were $f=f_\mathrm{p}$, $f=0.6f_\mathrm{p}$ and $f=1.25f_\mathrm{p}$, respectively. One should note that in the last simulation (Fig. 10(c)), the external magnetic field was reversed (see the top diagram of Fig. 10(c)) to support forward-propagating (+x) EM wave. The lossy InSb with $\nu=0.001$ were considered in the FEM simulations to investigate the loss effect on the one-way INZ modes. It is obvious that the propagation properties of the launched EM waves in the simulations are similar to those in the simulations of microwave POBs demonstrated in the last subsections. Moreover, the phase angles remain the same in the thin parts of the three structures, which are in agreement with the calculation results illustrated in Fig. 9. Therefore, we conclude that the loss in the terahertz POBs will not significantly affect the propagation properties of the EM modes such as the INZ modes, and the similar result was demonstrated in the INZ metamaterials with cylindrical defects\cite{Nguyen:To}. Finally, the transmission efficiencies of the INZ modes in the microwave/terahertz POBs according to our simulations can be more than $85\%$. This high-performance characteristic is due to the 'topologically' unidirectional propagation properties in our design POBs\cite{Tsakmakidis:No}, whereas the supercoupling transmission suggested in other ENZ/INZ structures always require for ultra-small electrical scale(s)\cite{Silveirinha:Th,Silveirinha:Tu}.

	\section{Conclusion}
	We have investigated the tunable broadband index-near-zero (INZ) modes in simple MO heterostructures. As an example of the potential applications of the INZ modes, we further proposed the low-loss microwave perfect optical buffers (POBs) and terahertz POBs. In the microwave regime, we have demonstrated that by carefully modifying the thickness ($d_1$) of the dielectric layer, the INZ modes can be tuned in a broad band (i.e. the COWP band). The group velocities $v_\mathrm{g}$ of the INZ modes were found to be nonzero, and $v_\mathrm{g}$ decreased when decreasing $d_1$, indicating the slowed wave in the thin MO structures. Based on these results, the 2D and 3D microwave POBs have been designed, and the tunable broadband characteristic of the INZ modes have been confirmed in the numerical simulations.
	
	In addition, we have extended our research to the terahertz regime. By using the perfect magnetic conductor walls, three types of terahertz INZ mode-based POBs were proposed based on the PMC-dielectric-semiconductor-PEC (MDSE) and the PMC-dielectric-semiconductor-PMC (MDSM) configurations. Moreover, different from the microwave INZ modes, the terahertz INZ modes, according to our analyses, can be engineered by changing the thickness of the semiconductor in the MDSE structure and changing the thickness of the dielectric in the MDSM structure. We also have proposed a promising way to turn 'on' or 'off' the INZ region by easily varying the external magnetic field, making our MO materials-based INZ modes highly maneuverable in practice. Owing to the characteristics of one-way propagation, zero phase shift (ZPS), broad band and tunability, the microwave and terahertz INZ modes in the simple MO heterostructures are with great potential for high-performance and directional wireless communication, cloaking, holographic imaging and other optical functional devices in integrated optical communication besides POBs.

	\section*{Funding information}
	National Natural Science Foundation of China (NSFC) (61927813, 61865009); Funding of Southwest Medical University (20/00160222); Department of Science and Technology of Sichuan Province (14JC0153); Science and Technology Strategic Cooperation Programs of Luzhou Municipal People’s Government and Southwest Medical University (2019LZXNYDJ18);the European Union’s Horizon 2020 research and innovation program (H2020, no.713683) under the Marie Sklodowska-Curie grant agreement no. 713683 (COFUNDfellowsDTU); the Spring Buds Program from NIMTE CAS starting grant.

	
	\bibliographystyle{unsrt}
	\bibliography{mybib}
	

\end{document}